\documentclass[a4paper, amsfonts, amssymb, amsmath, reprint, showkeys, nofootinbib, twoside]{revtex4-1}
\usepackage[english]{babel}
\usepackage{bm}
\usepackage[utf8]{inputenc}
\usepackage[colorinlistoftodos, color=green!40, prependcaption]{todonotes}
\usepackage{amssymb}
\usepackage{amsthm}
\usepackage{mathtools}
\usepackage{physics}
\usepackage{xcolor}
\usepackage{graphicx}
\usepackage[left=23mm,right=13mm,top=35mm,columnsep=15pt]{geometry} 
\usepackage{adjustbox}
\usepackage{placeins}
\usepackage[T1]{fontenc}
\usepackage{lipsum}
\usepackage{csquotes}
\usepackage[pdftex, pdftitle={Article}, pdfauthor={Author}]{hyperref} 
\begin{document}

\title{Reconstructing the redshift evolution of Type Ia supernovae absolute magnitude}

\author{Rodrigo von Marttens}
\email{rodrigomarttens@ufba.br}
\affiliation{Universidade Federal da Bahia, Salvador-BA, 40170-110, Brasil}
\affiliation{PPGCosmo, Universidade Federal do Espírito Santo, Vitória-ES, 29075-910, Brasil}

\author{Javier Gonzalez}
\email{javiergonzalezs@academico.ufs.br}
\affiliation{Universidade Federal de Sergipe, São Cristóvão - SE, 49107-230, Brasil}

\author{Jailson Alcaniz}
\email{alcaniz@on.br}
\affiliation{Observatório Nacional, Rio de Janeiro - RJ, 20921-400, Brasil}

\date{\today} 

\begin{abstract}
This work investigates a potential time dependence of the absolute magnitude of Type Ia Supernovae (SN Ia). Employing the Gaussian Process approach, we obtain the SN Ia absolute magnitude and its derivative as a function of redshift. The data set considered in the analysis comprises measurements of apparent magnitude from SN Ia, Hubble rate from cosmic chronometers, and the ratio between angular and radial distances from Large-Scale Structure data (BAO and voids). Our findings reveal good compatibility between the reconstructed SN Ia absolute magnitudes and a constant value. However, the mean value obtained from the Gaussian Process reconstruction is $M=-19.456\pm 0.059$, which is $3.2\sigma$ apart from local measurements by Pantheon+SH0ES. This incompatibility may be directly associated to the $\Lambda$CDM model and local data, as it does not appear in either model-dependent or model-independent estimates of the absolute magnitude based on early universe data. Furthermore, we assess the implications of a variable $M$ within the context of modified gravity theories. Considering the local estimate of the absolute magnitude, we find $\sim3\sigma$ tension supporting departures from General Relativity in analyzing scenarios involving modified gravity theories with variations in Planck mass through Newton's constant.
\end{abstract}


\maketitle

\section{Introduction \label{sec:intro}}

One of the most intriguing discoveries of modern cosmology is that our Universe is undergoing a phase of accelerated expansion. Within the framework of Einstein's General Relativity (GR) theory and the Cosmological Principle (CP), such a phenomenon can be explained from an enigmatic component known as dark energy, characterized by a negative pressure. The standard theoretical description of dark energy relies on the cosmological constant, denoted by $\Lambda$, which is analogous to a fluid featuring constant energy density and pressure, with EoS given by $p_{\Lambda}=-\rho_{\Lambda}$.

The first direct observational evidence for the accelerated expansion of the Universe emerged through the analysis of Type Ia Supernovae (SNe Ia) data~\cite{SupernovaSearchTeam:1998fmf,SupernovaCosmologyProject:1998vns}, which profoundly altered our cosmic understanding~\cite{Durrer:2007re}. Subsequently, the SNe Ia data have become essential for understanding the late-time dynamics of the Universe~\cite{SDSS:2014iwm,Pan-STARRS1:2017jku,Brout:2022vxf}. Since then, other observational results have confirmed the accelerated cosmic expansion~\cite{Planck:2018vyg,DES:2021wwk,Ivanov:2020ril}. 

With respect to SN Ia, these events mark the culmination of the accretion/merger process involving a carbon-oxygen white dwarf within a binary system. In a simplified sense, when a white dwarf approaches the Chandrasekhar limit of about $1.4\ M_{\odot}$, the electron degeneracy pressure becomes insufficient to counteract gravitational collapse~\cite{shapiro1983physics}. The consequence of this instability is a cataclysmic explosion, radiating luminosity equivalent to that of an entire galaxy. This well-established astrophysical mechanism suggests that SN Ia may possess an intrinsic (constant) absolute magnitude $M$, making them suitable as ``standardizable'' candles. In such a scenario, the apparent magnitude $m$ and the absolute magnitude $M$ are related via
\begin{equation} \label{eq:mu}
    \mu\equiv m-M=5\log_{10}\left[\frac{D_{L}\left(z\right)}{1\,{\rm Mpc}}\right]+25\,,
\end{equation}
where $\mu$ is the distance modulus and $D_{L}\left(z\right)$ is the luminosity distance. Considering a spatially flat universe, which is assumed in this work, the luminosity distance can be written as
\begin{equation} \label{eq:Dl}
    D_{L}\left(z\right)=c\left(1+z\right)\int_0^z\frac{d\tilde{z}}{H\left(\tilde{z}\right)}\,,
\end{equation}
where $H\left(z\right)$ is the Hubble rate.

Observations of SN Ia provide measurements of the apparent magnitude across a range of redshifts. The conventional approach to parameter selection using SN Ia data involves adopting a specific cosmological model, explicitly defining the luminosity distance (i.e., the Hubble rate along the redshift), and treating the absolute magnitude as a nuisance parameter, assumedly constant for the entire dataset\footnote{It is noteworthy that an additional parameter $\Delta M$, contingent on the host galaxy mass, can also be considered (see~\cite{SDSS:2014iwm} for further details).}. Furthermore, guided by the foundational assumption that SN Ia acts as a standardizable candle (an essential hypothesis for their use as a cosmological probe), the analysis also incorporates the light curve parameters as additional nuisances~\cite{SNLS:2007cqk,Kenworthy:2021azy}. In particular, the role of the absolute magnitude in parameter selection with SN Ia data has recently been a subject of extensive discussion in the literature~\cite{Camarena:2019moy,Camarena:2021jlr}. Moreover, due to the degeneracy between the Hubble constant and the absolute magnitude for parameter estimation with SN Ia data, determining $M$ is crucial in the $H_0$-tension debate.

In contrast to the conventional assumption treating the absolute magnitude of SN Ia as a constant, recent studies have questioned this paradigm, suggesting the potential for a variable absolute magnitude of SN Ia influenced by astrophysical factors~\cite{Kang:2019azh,Kim:2019npy,Rigault:2014kaa}. In this context, this work explores the feasibility of a time-dependent absolute magnitude  (redshift-dependent) $M\left(z\right)$. To this end, we employ the well-known non-parametric method, namely Gaussian Process, to obtain the SN Ia absolute magnitude as a function of redshift directly from the data. Differently from the approach commonly used in the literature, we assume  the CPL parameterization as the prior mean functionand  marginalise over the hyperparameters of the kernel and the prior  as discussed below (see Ref. \cite{Hwang:2022hla}). It is worth mentioning that other different analysis for assessing a possible time dependency of the SN Ia have already been proposed, for example, considering specific time-dependent parameterizations for $M\left(z\right)$~\cite{Sapone:2020wwz,Kumar:2021djt} or considering a binning in different redshift values~\cite{Koo:2020ssl}. 

Specifically, our methodology involves the reconstruction of the SN Ia absolute magnitude through observational data from three different cosmological observables: (i) SN Ia apparent magnitude; (ii) Large Scale Structure (LSS) data from galaxy-galaxy and void-galaxy clustering; (iii) Cosmic Chronometers (CC) data. Our approach is best understood as a null test, with the null hypothesis being that the SN Ia absolute magnitude is constant. This approach offers the advantage of enabling an investigation into deviations from the null hypothesis without being restricted to a specific model. Many distinct null tests have already been introduced to explore deviations from the standard cosmological model~\cite{vonMarttens:2018bvz,Andrade:2021njl} (for a comprehensive view, see~\cite{Euclid:2021frk} for a Euclid forecast encompassing key null tests from the existing literature). Using ML tools to obtain a time evolution of the observables also keeps us from relying on direct cosmological and/or astrophysical assumptions.

This work is organized as follows: In Sec. \ref{sec:theory}, we present the mathematical relations between the cosmological observables used to calculate the redshift evolution of the absolute magnitude,$M(z)$. In Sec. \ref{sec:ml}, we describe the data and discuss the  probability framework of GPs as a regression method. In Sec \ref{sec:results}, we present the GP reconstructions of the observables considered, the calculated redshift evolution of $M$ and its implications for the $H_0$ tension and modified gravity. Finally, we conclude in Sec. \ref{sec:conclusions}.

\section{Theory \label{sec:theory}}

In this section, we establish the theoretical foundation of our approach to evaluate whether the absolute magnitude of SN Ia remains constant over time. We begin with Eq.~\eqref{eq:mu}, which can be rearranged to express $M$ in terms of the apparent magnitude and the luminosity distance, as follows:
\begin{equation} \label{eq:Mz}
    M=m-5\log_{10}\left[\frac{D_{L}\left(z\right)}{1\,{\rm Mpc}}\right]-25\,.
\end{equation}

As previously mentioned, our approach uses diverse observational datasets to reconstruct the individual components on the right-hand side of Eq.\eqref{eq:Mz}. To begin with, for the apparent magnitude $m$, we  directly employ the SN Ia data from the Pantheon+SH0ES dataset\cite{Scolnic:2021amr}. We chose the Pantheon+SH0ES dataset over more recent samples (e.g., Union 3 [Ref. \cite{2023arXiv231112098R}] and DES Y5 [Ref. \cite{Abbott:2024agi}]) because it allows us to break the degeneracy between $H_0$ and $M$.

As for the second term on the right-hand side, dealing with the luminosity distance we find convenient to apply the Cosmic Distance Duality Relation (CDDR)\footnote{In recent years, several analyses have observationally tested this relation and verified its validity within $\gtrsim 2\sigma$ (see e.g. \cite{Goncalves:2019xtc,Holanda:2012at} and references therein).}, which is expressed as,
\begin{equation} \label{eq:CDDR}
    D_{L}\left(z\right)=\left(1+z\right)^{2}D_{A}\left(z\right)\,,
\end{equation}
where $D_{A}\left(z\right)$ is the angular diameter distance, which can be achieved from LSS data. Combining Eqs.~\eqref{eq:Mz} and~\eqref{eq:CDDR}, we obtain,
\begin{equation} \label{eq:M}
    M=m-5\log_{10}\left[\left(1+z\right)^{2}\frac{D_{A}\left(z\right)}{1\,{\rm Mpc}}\right]-25\,.
\end{equation}
In practice, LSS data alone do not directly yield the angular diameter distance $D_A(z)$. Instead, it often provides a combination with another quantity. For instance, in~\cite{eBOSS:2020yzd}, you can find measurements of $D_M(z)/r_d$ derived from Baryon Acoustic Oscillations (BAO). Here, $D_M(z)$ stands for the comoving angular diameter distance, defined as $\left(1+z\right)D_A(z)$, and $r_d$ corresponds to the sound horizon at the drag epoch. In our work, in addition to  BAO data, we incorporate data derived from the cross-correlation between galaxy and void positions~\cite{Woodfinden:2023oca}. In both cases (BAO and voids), the measurable quantity is the ratio $D_M(z)/D_H(z)$, where $D_H(z)$ is defined by $D_H(z) = c/H(z)$. Consequently, Eq.~\eqref{eq:M} can be reformulated as follows:
\begin{equation} \label{eq:MDH}
    \small
    M=m-5\log_{10}\left[\frac{D_{M}\left(z\right)}{D_{H}\left(z\right)}\right]-5\log_{10}\left[\left(1+z\right)\frac{D_{H}\left(z\right)}{1\,{\rm Mpc}}\right]-25\,,
\end{equation}
where $D_{H}(z)$ is introduced by dividing and multiplying it by the expression within the logarithm in Eq.\eqref{eq:M}. The second term on the right-hand side of Eq.\eqref{eq:MDH} can then be reconstructed directly from the LSS data, incorporating information from Baryon Acoustic Oscillations (BAO) and voids. In contrast, the third term on the right-hand side of Eq.~\eqref{eq:MDH} relies solely on the Hubble rate, allowing for its reconstruction using data from Cosmic Chronometers. Expanding the Hubble rate, Eq.~\eqref{eq:MDH} assumes the following form:
\begin{equation} \label{eq:ME}
    \small 
    M=m-5\log_{10}\left[\frac{D_{M}\left(z\right)}{D_{H}\left(z\right)}\right]-5\log_{10}\left[\frac{1+z}{H\left(z\right)}\left(\frac{c}{1\,{\rm Mpc}}\right)\right]-25\,.
\end{equation}

Ultimately, Eq.~\eqref{eq:ME} is our main relation. It delivers the $M$ in terms of components that can be obtained directly from the observational data in a model-independent way.

\section{Nonparametric statistics \label{sec:ml}}

This section provides an in-depth discussion of the data sources and nonparametric methodology applied in this work. We begin by presenting a comprehensive overview of the data sets and how they are applied in reconstructing each term on the right-hand side of Eq.~\eqref{eq:ME}. Subsequently, we describe the GP method employed for the reconstruction approach.

\subsection{The data \label{sec:data}}

As the data, we employ the latest publicly available background data, comprising:

\begin{itemize}
    \item \textbf{Type Ia Supernovae: }For the determination of the apparent magnitude as a function of redshift, we make use of the Pantheon+SH0ES dataset~\cite{Scolnic:2021amr}. The Pantheon+ catalog comprises 1,701 light curves sourced from 1,550 distinct Type Ia Supernovae (SN Ia), encompassing a redshift range from $z=0.001$ to $2.26$~\cite{Brout:2022vxf}. It represents the most up-to-date and comprehensive repository of publicly accessible SN Ia data. Additionally, the SH0ES dataset~\cite{Riess:2021jrx} provides us with a unique opportunity, enabling the utilization of 77 objects as calibrators. The distances of these objects,  obtained through geometric methods, can be directly employed without relying on a specific cosmological model. Employing these calibrators enables the simultaneous constraints on both $M$ and $H_0$ solely using SN Ia data. 

    \item \textbf{LSS (BAO + voids): }The large-scale structure (LSS) of the Universe holds a wealth of information concerning the background and perturbative dynamics of the universe. In this work, we use information from two distinct sources within the LSS: ($i$) the baryon acoustic oscillation (BAO), a feature remnant of primordial sound waves employed as a standard ruler, and ($ii$) data derived from the distribution of galaxies surrounding cosmic voids.

    In regard to BAO, observations conducted along the line of sight and perpendicular to it provide measurements of $D_H\left(z\right)/r_d$ and $D_M\left(z\right)/r_d$ respectively. Here, $r_d$ represents the comoving sound horizon, and the expressions for $D_H\left(z\right)$ and $D_M\left(z\right)$ are as follows,
    \begin{eqnarray}
        D_H\left(z\right)&=&\frac{c}{H\left(z\right)}\,, \label{eq:DH} \\
        D_M\left(z\right)&=&\int_0^{z}\frac{c}{H\left(z'\right)}dz'\,. \label{eq:DM}
    \end{eqnarray}
    Note that a spatially flat universe is assumed in Eq.~\eqref{eq:DM}. 

    Real observations often generate anisotropic BAO measurements, potentially failing to accurately depict the real galaxy map. In such cases, restoring isotropy in the map can be achieved through the Alcock-Paczynski test. This test aims to ascertain the correct ratio between $D_M\left(z\right)$ and $D_H\left(z\right)$ (the Alcock-Paczynski parameter) when converting redshift to distance, to reinstate isotropy.

    On the other hand, void-galaxy cross-correlations lack a characteristic distance scale suitable for direct use as a standard ruler. Nevertheless, under the assumption of a statistically isotropic Universe, the arrangement of galaxies within and around voids is expected to result in an observable distribution demonstrating spherical symmetry. The expected observation of this spherically symmetrical pattern is expected also upon the correct determination of the Alcock-Paczynski parameter. 
    
    In this work, we utilize the $D_{M}\left(z\right)/D_{H}\left(z\right)$ data sourced from both BAO and voids, as outlined in Tab.~1 of Ref.~\cite{Woodfinden:2023oca}. It is important to mention that, even though LSS data uses a fiducial cosmology, recent work has shown that this effect can be considered negligible~\cite{Carter:2019ulk,Pan:2023zgb,Sanz-Wuhl:2024uvi}.

    \item \textbf{Cosmic Chronometers: }Cosmic chronometers refer to obtain the differential age of galaxies that undergo passive evolution, with known redshifts. In this approach, the derivative of cosmic time with respect to redshift is approximated by the ratio of the variation in the age of the galaxy with redshift\footnote{This approximation is reasonable under the assumption that the examined galaxies were formed at the same epoch.}. Using this approximation, the Hubble rate can be directly inferred as,
    \begin{equation} \label{eq:Hz}
        H\left(z\right)=-\frac{dz}{dt}\frac{1}{1+z}\,.
    \end{equation}

    In this work, we utilize the publicly available data from~\cite{Moresco:2016mzx}. 
\end{itemize}

With this dataset, we possess all the necessary components to derive all terms on the right-hand side of Eq. ~\eqref{eq:ME}, thereby enabling the determination of the absolute magnitude of SN Ia as a function of redshift.

\subsection{ Gaussian Processes \label{sec:gp}}

We use a non-parametric approach, the so-called Gaussian Process (GP) regression, to reconstruct the behaviour of the SN Ia apparent magnitude, the Hubble rate and the $D_M/D_H$ as functions of redshift. Gaussian Process is a method of the spatial statistics that extends a random Gaussian variable to a random Gaussian function. In this approach, the observations correspond to a realization of a random function which values in different domain points are correlated. This correlation is described by a kernel function, which is defined from the expected properties of the analyzed observable such as differentiability and independence between distant points. For this reason, a quadratic exponential kernel is usually selected, as done in this work. 

Moreover, GP analysis is based on Bayesian statistics. Therefore, a crucial aspect lies in the selection of a prior mean function to describe the data behaviour, and the resulting regression is derived from the conditional probability of this function given the data.
Following the Ref.~\cite{Hwang:2022hla}, we opt to use the CPL model solution \cite{Chevallier:2000qy,Linder:2002et} as a prior mean function instead of using a zero mean function. The motivations about this choice are elaborated upon in Sec.~\ref{sssec:gpmean}. It is worth mentioning that, for this choice, while the mean function plays an important role in determining the shape and the uncertainty of the cosmological observable, the cosmological parameters, being marginalised over, do not exert a direct influence.

The choice of the CPL description instead of the $\Lambda$CDM model is justified by the number of degrees of freedom related to the dark energy component. In the CPL model there are two extra dark energy parameters compared to the $\Lambda$CDM model, the $w_0$ and $w_a$ EoS parameters. The effect of these additional degrees of freedom in the CPL case will appear on the error propagation. From a physical point of view, this effect can be interpreted as if we are taking into account the potential degrees of freedom related to dark energy and their potential correlations. In this context the  analysis using the CPL parameterization can be considered more robust. 


\subsubsection{On the Gaussian Process mean function \label{sssec:gpmean}}

It is crucial to mention the ongoing debate in the literature concerning the appropriate application of the Gaussian Process method in assessing cosmological problems. A key point of the debate regards around the choice of the prior mean function. Given that the Gaussian Process methodology is independent of assumptions regarding the cosmological model, in order to avoid biased results, most works opt for a choice of a zero mean function (e.g., \cite{Seikel:2012uu,Gonzalez:2016lur,vonMarttens:2018bvz,vonMarttens:2020apn,Li:2015nta,Bengaly:2020vly,Li:2022cbk,Wu:2022fmr,Colaco:2023gzy,Jesus:2019nnk,Gonzalez:2017fra,Gonzalez:2024qjs}. In the case of reconstructing the distance modulus of SN Ia or equivalently the apparent magnitude, this approach results in oscillations which are evidently ``unphysical'' (e.g., see Fig. 2 in Ref. \cite{Hwang:2022hla} and results in Appendix \ref{sec:GP0}). In contrast, a recent work~\cite{Hwang:2022hla} suggests that the prior mean function should correspond to a more general function than the standard cosmological model, such as  CPL parameterization.  However, to mitigate any model-dependency or bias, the final analysis of the Gaussian Process  must be conducted after marginalising over the cosmological and kernel parameters.

As mentioned previously, in this work we adopt the second approach using the CPL model solution as the prior mean function, even though we also present an analysis employing the zero mean function in the Appendix~\ref{sec:GP0}. Our decision to align with the approach proposed in Ref.~\cite{Hwang:2022hla} is mainly motivated by the following reasons:
\begin{itemize}
    \item We consider that the oscillations resulting from selecting the zero mean function lack a physical basis.

    \item CPL model is widely acknowledged for its accurate fit to the background data utilized in this study and it is general enough to allow the flexibility of a non-parametric approach in a observable reconstruction. Consequently, it is reasonable to consider that the ``correct'' model would exhibit a similar characteristic shape. Essentially, employing the CPL model and subsequently marginalising over its parameters can be interpreted as a search for deviations from an effective model.

    \item Considering the zero mean function ignores prior knowledge about the behaviour of the apparent magnitude, which is a logarithmic function  of redshift for a type of objects with a defined intrinsic luminosity. Specifically, a zero function can not described the apparent magnitude divergence at $z=0$. 
On the other hand, in the case of employing a mean function based on a cosmological model and subsequently marginalising over the cosmological parameters, the errors associated with these parameters are propagated in the reconstruction, which is more appropriate for this observable. 
\end{itemize}

A further discussion about these different approaches is conducted in Ref.~\cite{Hwang:2022hla}. For the specific case analysed in this work, we study the impact of using the zero mean function in Appx.~\ref{sec:GP0}. 

\subsubsection{On the Gaussian Process parameter distribution estimation and reconstruction \label{sssec:gpdist}}

Based on the assumption that the data, $y$, and the function describing an observable, $f$, are components of the same multivariate Gaussian distribution (components of a GP), we can compute the function of the observable  given the data using the conditional probability. As stated above, the joint distribution is determined by a covariance matrix (kernel) and a prior mean function, $\mu$. We opt for the usual squared exponential kernel:

\begin{equation}
\label{SE}
    k(z,z')=\sigma^2\exp \left( -\frac{(z-z')^2}{2l^2}\right),
\end{equation}
where $\sigma$ and $l$ are the hyperparameters of the method, which are associated with the uncertainty and spatial variation of the function, respectively. In this context, $l$ determines the function's smoothness by defining the effective range within which data influences the prediction. Beyond this range,  data information is limited, and the function is determined by the prior mean. Thus, for small $l$ values, the reconstruction transitions from data-driven values to those given by the prior mean $\mu$, which can result in the appearance of wiggles. In regions where the influence ranges of different observations overlap, a smooth reconstruction is achieved.

 We emphasize that $\mu$ is a function of $z$ (the domain variable), but it may also depend on a set of parameters that we call $\bm \theta_\mu$.

The resulted reconstruction given the data and specific hyperparameter values, $p(\bm f(\bm z^*)|\bm y,\sigma,l,\bm \theta_\mu )$, is a multivariate Gaussian distribution with mean given by

\begin{equation}
\label{eq:meanrec}
     \bar{\bm f}(\bm z^*)=\bm \mu( \bm z^*)+\bm K(\bm z^*, \bm z)[\bm K(\bm z,\bm z)+ \bm C]^{-1}\bm (\bm y- \bm \mu(\bm z))^T 
\end{equation}
and covariance given by
\begin{equation*}
    \textbf{Cov}(\bm f(\bm z^*),\bm f(\bm z^*))=
\end{equation*}
\begin{equation}
\label{eq:covrec}
    \bm K(\bm z^*, \bm z^*)+\bm K(\bm z^*, \bm z)[\bm K(\bm z,\bm z)- \bm C]^{-1}\bm K(\bm z, \bm z^*),
\end{equation}
where  bold capital and bold lowercase letters represent matrices and vectors, respectively;  $\bm z^*$ represents the vector of redshifts where we want to reconstruct the observable and $\bm z$ the vector of  redshift data; $\bm C$ is the covariance matrix of the data (for uncorrelated data it is diagonal), and $\bm K$ is the assumed covariance between the $f$ values which matrix elements are given  by $[\bm K(\bm z,\bm z)]_{i,j}=k(z_i,z_j)$ (Eq. (\ref{SE})). 

To explore the hyperparameter distributions, the marginal likelihood is considered,

\begin{equation}
\label{eq:mlikelihood}
    \mathcal{L}= p(\bm y|\sigma, l, \bm \theta_\mu)=  \frac{\exp(-(\bm y-\bm \mu)\widetilde{\bm K}^{-1}(\bm y-\bm \mu)^T/2)}{(2\pi)^{n/2}|\widetilde{\bm K}|^{1/2}},
\end{equation}

\begin{figure*}
\centering
\includegraphics[width = 0.32\textwidth, trim={0 0.3cm 0 0.2cm},clip]{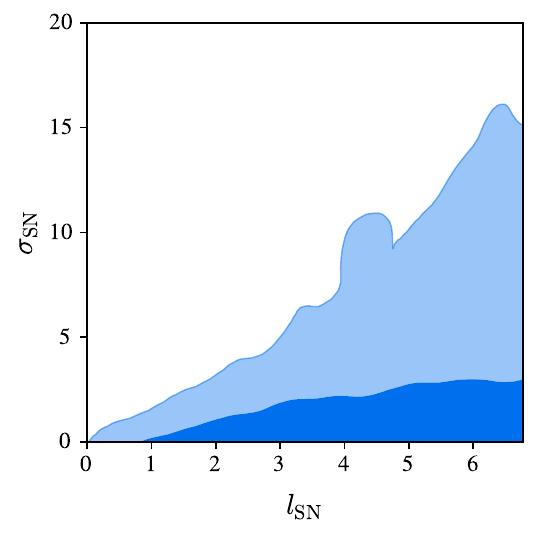}
\includegraphics[width = 0.32\textwidth, trim={0 0.3cm 0 0.2cm},clip]{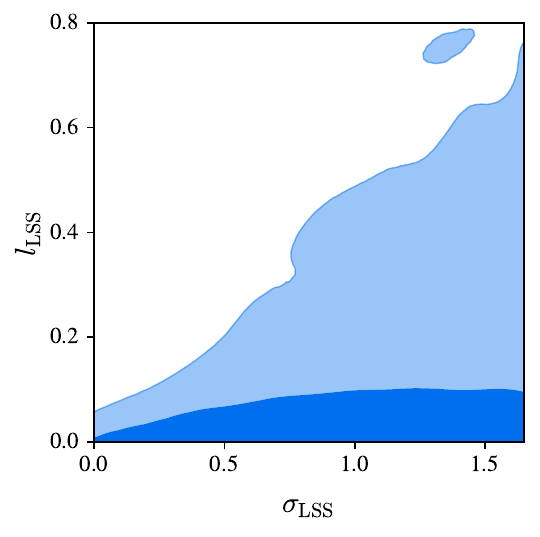}
\includegraphics[width = 0.32\textwidth, trim={0 0.3cm 0 0.2cm},clip]{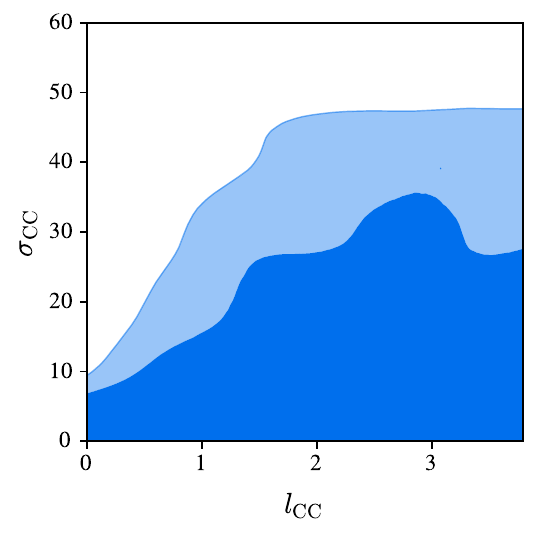}
\caption{Contour planes (2$\sigma$ confidence level) for the hyper-parameters of the GP analyses. \textbf{Left panel:} Results for the SN Ia analysis. \textbf{Middle panel:} Results for the LSS analysis. \textbf{Right panel:} Results for the CC analysis. \label{fig:GP_hyperparams}}
\end{figure*}

\noindent where  $\widetilde{\bm K}=\bm K(\bm z,\bm z)+\bm C$. By using the Bayes theorem and the marginal likelihood in Eq. (\ref{eq:mlikelihood}), we obtain the posterior distribution of the hyperparameters $\sigma$, $l$ and $\bm \theta_\mu$,
\begin{equation}
\label{eq:hyper_posterior}
    p(\sigma, l, \bm \theta_\mu|\bm y)\propto p(\bm y|\sigma, l, \bm \theta_\mu)\pi(\sigma, l, \bm \theta_\mu),
\end{equation}
being $\pi(\sigma, l, \bm \theta_\mu)$ the assumed prior distribution of the hyperparameters. As seen in Eqs. (\ref{eq:meanrec}) and (\ref{eq:mlikelihood}), the mean value of the reconstruction, as well as the marginal likelihood and consequently the hyperparameter distributions, are explicitly dependent on the prior mean function.  On the other hand, in Eq. (\ref{eq:covrec}), the covariance of the reconstruction appears to be independent of $\mu$. However, since the hyperparameter distributions depend on $\mu$, the covariance of the reconstruction implicitly also depends on it.

If the posterior distribution of the hyperparameters  is peaked, a good approximation is to use the hyperparamters that  maximise Eq. \eqref{eq:hyper_posterior}. In this case, for a fixed $\bm \theta_\mu$ values, the reconstruction is directly given by Eqs. \eqref{eq:meanrec} and \eqref{eq:covrec}. However, for a robust Bayesian  analysis, we reconstruct the observable by marginalising over the complete set of hyperparameters, including covariance and prior mean function parameters, as follows:
\begin{equation}
\label{eq:marginzalised_rec}
  p(\bm f(\bm z^*)|\bm y) =\int p(\bm f(\bm z^*)|\bm y,\sigma,l,\bm \theta_\mu )p(\sigma, l, \bm \theta_\mu|\bm y)d\sigma dl d\bm \theta_\mu.
\end{equation}
In this approach, the final GP-reconstructed observable and its error are the mean function and the standard deviation of the distribution \ref{eq:marginzalised_rec}, respectively. In addition, as the Eq. \eqref{eq:marginzalised_rec} corresponds to the joint multivariate distribution of $f$   at different $z^*$-values,  it is also possible to calculate the covariance of the reconstructed function along the redshift range considered. 

The marginalised distribution $p(\bm f(\bm z^*)|\bm y)$ is estimated by performing  Monte Carlo samplings of the distributions  inside the integral in Eq. \eqref{eq:marginzalised_rec} as follows:
\begin{itemize}
    \item    A sample of the hyperparameters with distribution $p(\sigma, l, \bm \theta_\mu|\bm y)$ is obtained in a MCMC process utilizing the likelihood in Eq. \eqref{eq:mlikelihood} and we consider uninformative flat priors, $\pi(\sigma, l, \bm \theta_\mu)$, for all parameters. 
    \item This distribution of  $p(\sigma, l, \bm \theta_\mu|\bm y)$ is used to perform  Monte Carlo sampling  for each set of values $\{\sigma,l,\bm \theta_\mu\}$, given the distribution $p(\bm f(\bm z^*)|\bm y,\sigma,l,\bm \theta_\mu )$, which is defined as   a multivariate  Gaussian distribution with mean and covariance given by Eqs. \eqref{eq:meanrec} and \eqref{eq:covrec}. This process yields a complete sample of the marginalised distribution in Eq. \eqref{eq:marginzalised_rec}.
    \item The mean values, standard deviations and covariances of $\bm f(\bm z^*)$ at different redshifts are calculated from  this sample of $p(\bm f(\bm z^*)|\bm y)$ .
\end{itemize}
For  comprehensive details about the statistical theory of GPs, refer to the following Refs.\cite{RasmussenW06, Seikel:2012uu}.

The set of hyperparameters related to the prior mean function depends on the specific model adopted as $\mu$. As mentioned above, the main results of this work consider the flat CPL model as $\mu$,  defining the $\mu$-hyperparameter set as $\bm \theta_\mu=\{\Omega_m,H_0,w_0,w_a,M\}$. In the case of the results presented in Appendix \ref{sec:GP0}, it is assumed $\mu=0$, therefore, there are not hyperparameters associated to $\mu$ and the GP reconstruction only depends on the covariance parameters, $\sigma$ and $l$.

For the reconstruction of the observables $D_M(z)/D_H(z)$ and $H(z)$, the $\mu$ functions in the marginal likelihood (Eq.~\eqref{eq:mlikelihood}) are simply the solutions of Eqs.~\eqref{eq:MDH}, \eqref{eq:DM} and \eqref{eq:Hz} for the CPL parameterisation. Following the Pantheon+ analysis to constrain cosmological parameters \cite{Brout:2022vxf}, we consider the 1658 $m(z)$ data with $z>0.01$ and the calibration objects  to reconstruct  this observable . For the 77 calibrators, in Eq.~\eqref{eq:mlikelihood} $\mu=M+\mu^{\rm Cepheid}$, being $\mu^{\rm Cepheid}$ the distance moduli of Cepheid variables used as calibrators in the Pantheon+SH0ES dataset. Meanwhile, for the rest of the data, the $\mu$ function in Eq.~\eqref{eq:mlikelihood} is the solution of $m(z)$ in Eq.~\eqref{eq:mu} for CPL parameterisation. As explained in Ref.~\cite{Brout:2022vxf}, this approach allows to break the $H_0-M$ degeneracy.

\section{Results \label{sec:results}}

Herein, we present the results derived from the data and methodology delineated in Sec.~\ref{sec:ml}. More precisely, we use GP in order to obtain the individual reconstructions for the observables of interest: $m\left(z\right)$, $D_{M}\left(z\right)/D_{H}\left(z\right)$, and $H\left(z\right)$. Our main findings involve the reconstruction of SN Ia absolute magnitude and its derivative concerning redshift. These results enable us to assess whether $M$ remains constant or exhibits variations over time. Furthermore, we also illustrate specific outcomes related to the marginalisation of hyperparameters.

\subsection{Gaussian Process reconstructions}

\begin{figure}
\centering
\includegraphics[width = \columnwidth,trim={0 1cm 0 0.2cm},clip]{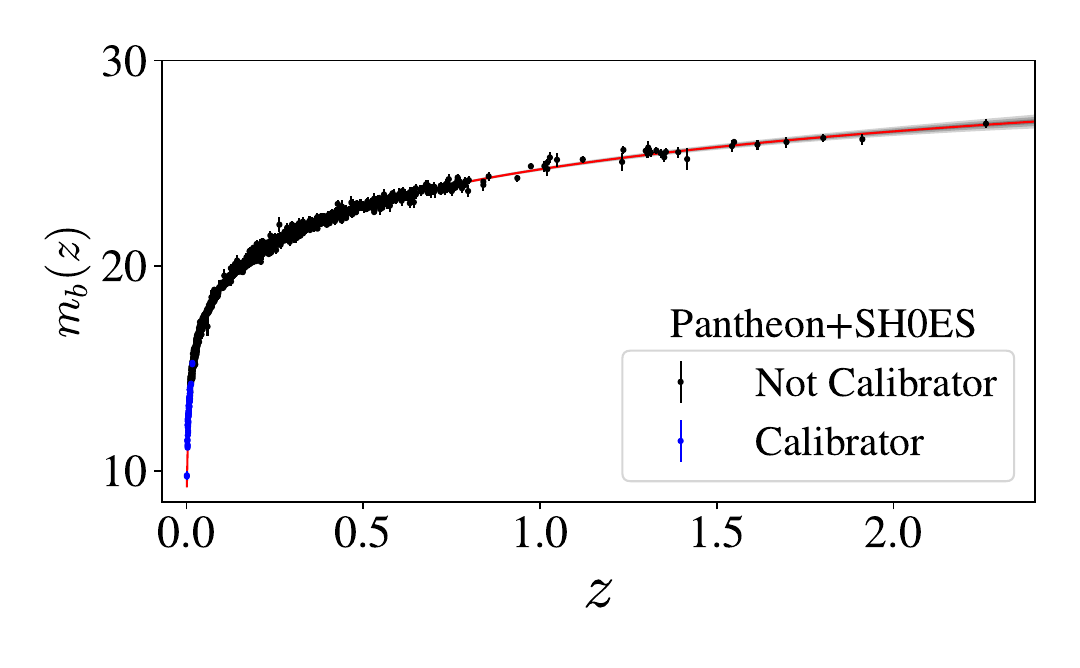}
\includegraphics[width = \columnwidth,trim={0 1cm 0 0.2cm},clip]{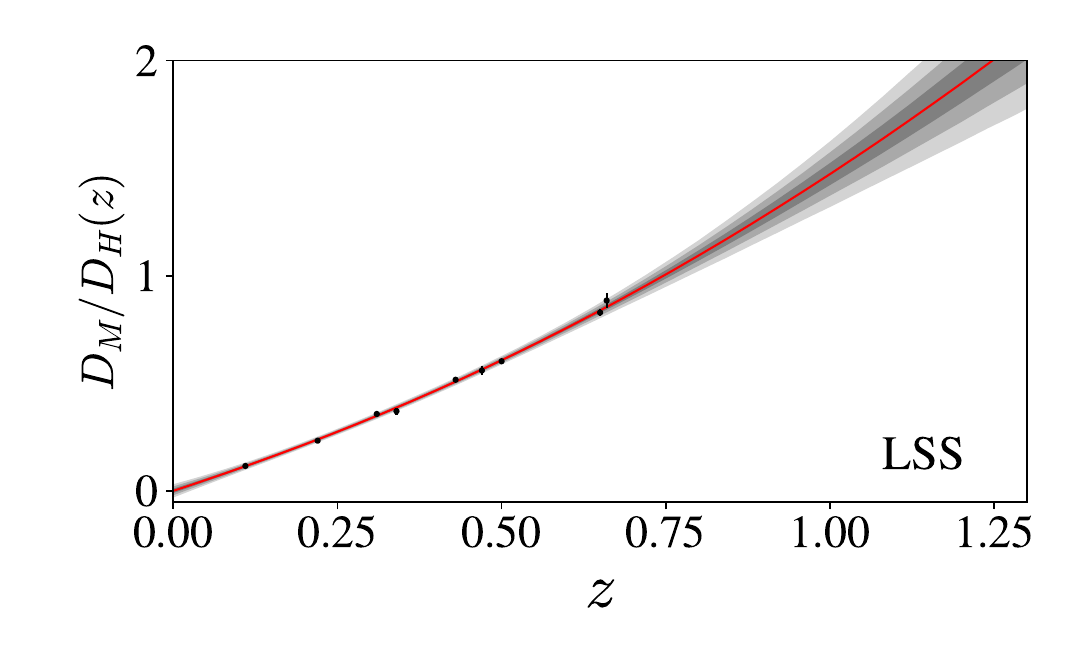}
\includegraphics[width = \columnwidth,trim={0 1cm 0 0.2cm},clip]{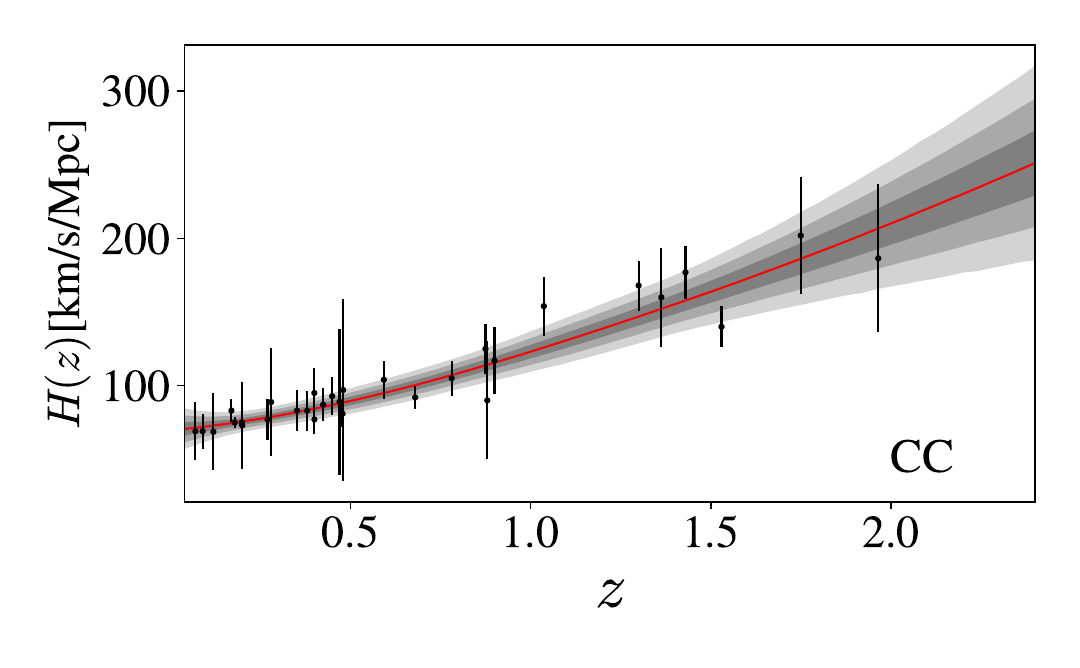}
\caption{Gaussian Process reconstructions for the isolated observables necessary to obtain the SN Ia absolute magnitude. \textbf{Top panel:} SN Ia aparent magnitude. \textbf{Middle panel:} Ratio between angular and radial distances. \textbf{Bottom panel:} Hubble rate. \label{fig:GP_rec}}
\end{figure}

This section is dedicated to presenting the results obtained using Gaussian Process. As discussed previously, it is worth remembering that we are using the solutions of the CPL model as the prior mean function in this analysis. Initially, in Fig.\ref{fig:GP_hyperparams}, we illustrate the contours concerning the hyperparameters $l$ and $\sigma$ for all observables. Notably, it is important to highlight that the findings regarding the hyperparameters in the SN Ia analysis align with the trends observed in the right panel of Fig. 3 in Ref.~\cite{Hwang:2022hla}. Specifically, our contours are slight bigger because we adopt the CPL model and marginalise over the cosmological parameters instead of $\Lambda$CDM as mean function with fixed parameter values. Furthermore, in our analysis we also consider the inclusion of $M$ as a free parameter. The SN hyperparameter probability distribution does not show preference for  small values of $l$ ($z<1$) compared with the highest  redshift difference of the data ($\Delta z \approx 2.26$). This fact implies a smooth reconstructed function for $m(z)$.

The individual reconstructions of observables $m(z)$, $D_M\left(z\right)/D_H\left(z\right)$, and $H\left(z\right)$ are depicted in Fig.~\ref{fig:GP_rec}. In each panel, the  solid red line denotes the expected function obtained from the analysis utilizing GPs, while the shaded grey regions indicate the corresponding confidence intervals at 1$\sigma$, 2$\sigma$, and 3$\sigma$ confidence levels. In addition to the mean and standard deviation values,  the covariances of each observable in different redshift are calculated as described in the previous section.

\begin{figure*}
\centering
\includegraphics[width = 0.49\textwidth, trim={0 0.3cm 0 0.2cm},clip]{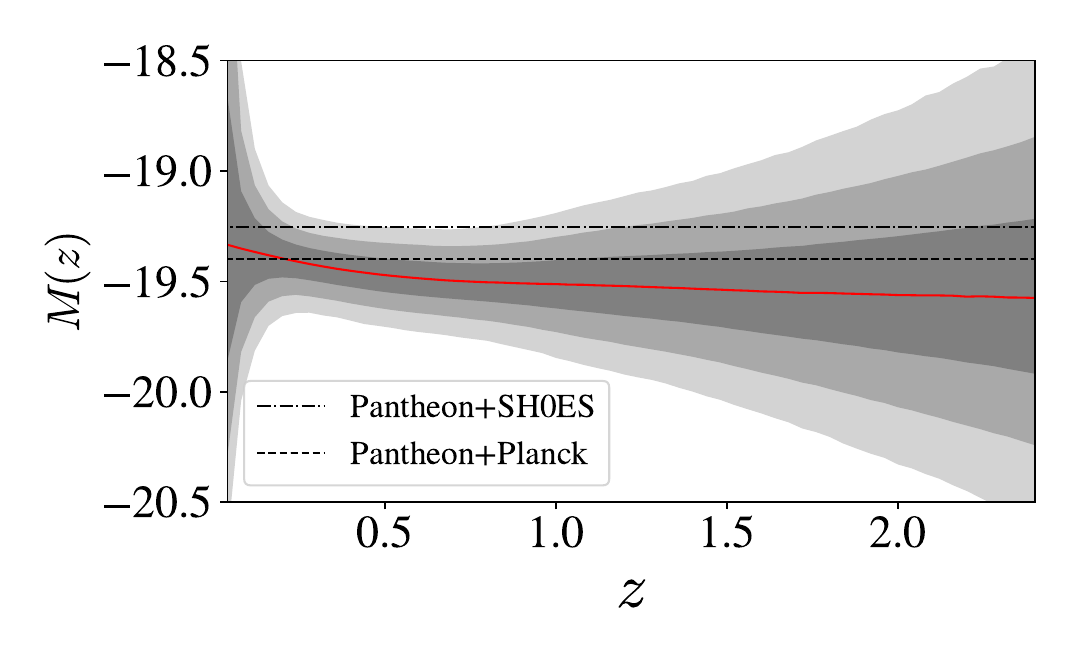}
\includegraphics[width = 0.49\textwidth, trim={0 0.3cm 0 0.2cm},clip]{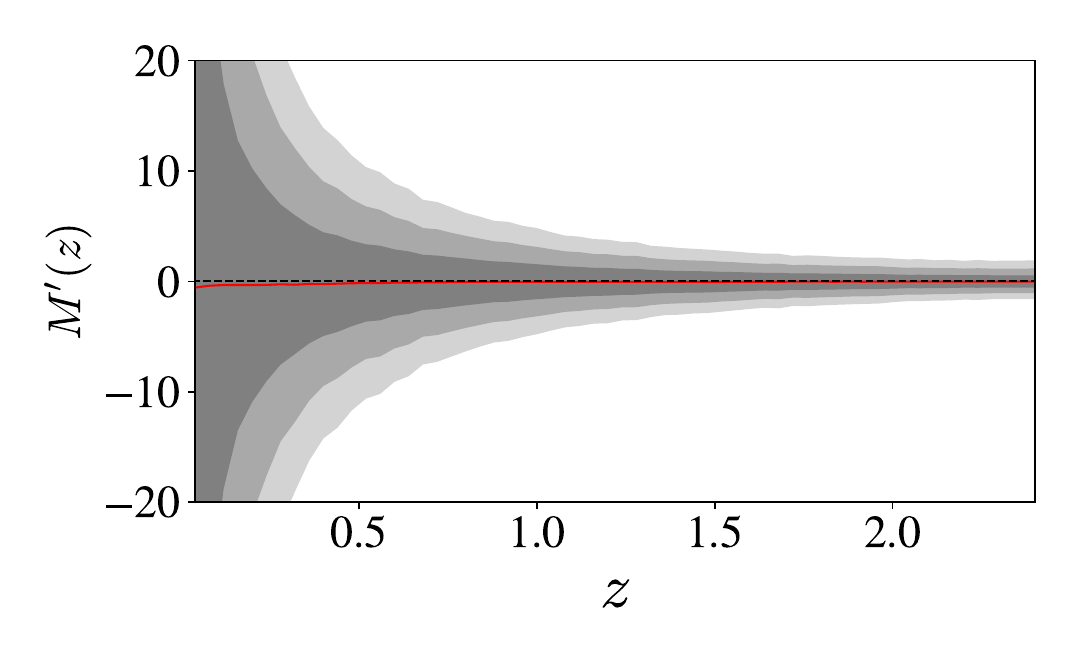}
\caption{\textbf{Left panel:} Absolute magnitude of SN Ia as a function of redshift. \textbf{Right panel:} Derivative of absolute magnitude of SN Ia as a function of redshift. The red lines and the shaded regions represent the mean value and the 1$\sigma$, 2$\sigma$ and 3$\sigma$ confidence levels, respectively. The dashed lines represent the estimates from the corresponding datasets. \label{fig:M_GP}}
\end{figure*}

The reconstruction of the apparent magnitude of supernovae is shown in the top panel of Fig.~\ref{fig:GP_rec}. Notably, it exhibits significant constraints up to redshift 1.5, beyond which the error bars escalate. This phenomenon is due to the fact that they are not homogeneously distributed in relation to the redshift. More precisely, less than 1\% of the data extends beyond $z > 1.5$. Regarding the LSS data, the reconstruction of the ratio between $D_M\left(z\right)$ and $D_H\left(z\right)$ is shown in the middle panel of Fig.~\ref{fig:GP_rec}. In this case, the observational data is highly precise and confined to $z\approx 0.7$, resulting in a pronounced restriction in reconstruction accuracy up to this threshold. Beyond this point, a notable increase in error bars is observed. In this context, the CPL mean function plays an important role in determining the shape of the extrapolation beyond $z > 0.7$. Finally, in the bottom panel of Fig.~\ref{fig:GP_rec}, the reconstruction of the Hubble parameter utilizing data from cosmic chronometers is presented. Here, it is clear that the reconstruction error bars are generally larger, due to the bigger uncertainties associated with cosmic chronometer data. Furthermore, the reconstruction constraint tends to weaken for higher redshifts, where the data is less densely populated.

With all individual reconstructions necessary to determine the absolute magnitude of SN Ia available, they can be combined in a Monte Carlo sampling process according to Eq.\eqref{eq:ME} to derive its time evolution. The result is depicted in Fig.~\ref{fig:M_GP}. Similar to the individual reconstructions, the solid red line denotes the mean function obtained from the GP analysis, while the gray regions denote the regions with $1\sigma$, $2\sigma$, and $3\sigma$ confidence levels. Moreover, Fig.~\ref{fig:M_GP} also includes two horizontal lines indicating the values of the SN Ia absolute magnitude obtained from local measurements  (dot-dashed) and the value derived from the analysis combined with CMB data (dashed). Examining Fig.~\ref{fig:M_GP}, it becomes evident that the time evolution of the absolute magnitude of SN Ia obtained with GP is notably consistent with a constant value. We also show in Fig.~\ref{fig:M_GP} the result obtained. 

However, it is noteworthy that the obtained result exhibits stronger alignment with the value derived from the analysis using CMB data. Within the interval $0.4 < z < 0.9$, the result obtained from local measurements approaches the 3$\sigma$ confidence level region. 
%

%

Finally, to further validate the result obtained for the SN Ia absolute magnitude, we also calculate its derivative with respect to the redshift by performing a Monte Carlo sampling, taking into account the correlations between the reconstructed  observables and their derivatives. The obtained result is depicted in Fig.~\ref{fig:M_GP}. In this scenario, regardless of the value of $M$, a zero derivative confirms the null hypothesis, i.e., a constant absolute magnitude of SN Ia. Once more, the solid red line represents the mean function obtained from the GP analysis, while the gray regions denote the regions with $1\sigma$, $2\sigma$, and $3\sigma$ confidence levels. As depicted in Fig.~\ref{fig:M_GP}, the GP reconstruction strongly aligns with a null derivative for the absolute magnitude of SN Ia.

Since the result corroborates with a constant, the values at different redshifts correspond to the same quantity $M$. For this reason,  we condense the best estimate of the absolute magnitude by taking the weighted average over the entire redshift range of the reconstructed function.

We  compute the weighted average over redshift of the  expected value  of $\bar{M}$  and its 1$\sigma$ confidence level  (red line and dark grey region in Fig. \ref{fig:M_GP}, respectively) as an estimate of the absolute magnitude. For a Gaussian Process, i.e., for continuous mean and covariance, this average is calculated with the expression:
\begin{equation}
    M= \frac{\int \bar{M}(z) C^{-1}_M(z,z')dzdz'}{\sqrt{\int C^{-1}_M(z,z')dzdz'}}   \pm \frac{1}{\sqrt{\int C^{-1}_M(z,z')dzdz'}},
\end{equation}
where $\bar{M}$ and $C^{-1}_M$ correspond to the mean and the inverse of the covariance of the reconstructed absolute magnitude, respectively. However, in practice, it reduces to the weighted average for correlated Gaussian measurements,

\begin{equation}
    M= \frac{\sum_{i,j} \bar{M}(z_i) C_M^{-1}(z_i,z'_j)}{\sqrt{\sum_{i,j} C_M^{-1}(z_i,z'_j)}}     \pm \frac{1}{\sqrt{\sum_{i,j} C_M^{-1}(z_i,z'_j)}}.
\end{equation}
 Finally, our best estimate for the absolute magnitude is $M=-19.456\pm 0.059$.

\subsection{Implications for $\Lambda$CDM model and the Hubble tension \label{sec:implicationsH0}}

\begin{figure}
\centering
\includegraphics[width = \columnwidth]{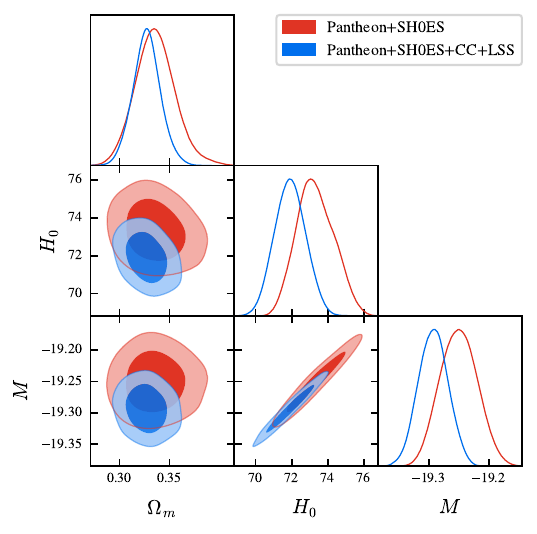}
\caption{Cosmological parameter inference assuming $\Lambda$CDM model.} 
\label{fig:triangle_LCDM}
\end{figure}

The results presented above provide evidence for  no evolution in the Absolute Magnitude of SN Ia. However, the inferred value is in 3.2$\sigma$ tension with the local estimate from Cepheid variables, $M=-19.243\pm 0.030$, and in agreement with the Pantheon+Planck estimate that assumes the $\Lambda$CDM model,  $M=-19.438\pm 0.007$ \cite{2023arXiv230702434C}. Since our estimate incorporates the SH0ES Cepheid information,  tension with the local value is not expected, but rather with the inference from Planck data, which implies that our result is driven by data rather than priors.

As the $H_0$ inference from CC is in total agreement with the $H_0$ from Planck, one possible explanation for this result could be the effect of CC data on the $H_0$ value, shifting $M$ towards the Pantheon+Planck estimate.  However, the $M$ inference being overwhelmed by the CC data is not a complete explanation. As shown in Table. \ref{table:results_LCDM} and Fig. \ref{fig:triangle_LCDM},  the $\Lambda$CDM model-dependent analyses of Pantheon+SH0ES+CC and Pantheon+SH0ES+CC+LSS data yield  $M$ estimates that are essentially the same than Pantheon+SH0ES, with no significant shift, which shows that CC data do not overwhelm over the SH0ES data.

Therefore, the found tension with local value appears to arise from the more flexible, model-independent approach used in this work. As discussed, the absolute magnitude depends on the statistical method employed (model-dependent or model-independent) when it is considered Cepheid variable data, but is this still valid when we consider early Universe data? In Ref. \cite{2023arXiv230702434C}, by performing a joint analysis of Pantheon+, BAO and CMB data, a model-independent estimate of the absolute magnitude is presented, $M=-19.401\pm 0.027$,  which  maintains agreement with the model-dependent estimate with early Universe data and with our result, and is in  3.4$\sigma$ tension with the local estimate.  

This has significant consequences for interpreting our results. The model-independent estimate of  $M$  using late-time Universe data aligns with both the model-dependent and model-independent values derived from including early Universe data. The significant tension appears when considering the $\Lambda$CDM  estimate based on local data, which could be interpreted as a $\Lambda$CDM  cosmology tension in the late-time Universe. 

\begin{table}[]
\begin{center}
\begin{tabular}{l | c c c}
 \hline\hline \\ [-2.0ex]
  & $\Omega_m$ & $H_0$ & M  \\ [0.5ex] 
 \hline \\ [-2.0ex]
  CC & $0.326_{- 0.055}^{+0.066}$& $67.8_{- 3.1}^{+2.9}$  & -\\ [1.0ex]
 \hline \\ [-2.0ex]
 SN & $0.333_{- 0.018}^{+0.017}$ & $73.6_{- 1.0}^{+1.1}$ &  $-19.243_{-0.030}^{+0.030}$  \\ [1.0ex]
 \hline \\ [-2.0ex]
 SN+CC & $0.313_{- 0.015}^{+0.016}$ & $72.2_{- 0.8}^{+0.9}$ & $-19.289_{- 0.024}^{+0.025}$  \\ [1.0ex]
 \hline \\ [-2.0ex]
 SN+CC+LSS & $0.327_{- 0.014}^{+0.013}$ & $71.9_{- 0.8}^{+0.9}$ & $-19.293_{- 0.025}^{+0.025}$  \\
 \hline\hline 
\end{tabular}
 \caption{Results of the statistical analysis for the $\Lambda$CDM model.}
\label{table:results_LCDM}
\end{center}
\end{table}

\subsection{Implications for modified gravity \label{sec:implications}}

An important class of modified gravity theories involves the replacement of Newton's constant with an effective gravitational term, which in the general case may exhibit time and scale dependencies [refs.],
\begin{equation}
    G_{\rm N}\longrightarrow G_{\rm eff}\left(z,k\right)\,.
\end{equation}
This replacement leads to changes impacting both the dynamics of the cosmological background and the perturbations. Besides cosmological implications, a time-dependent gravitational constant also influences astrophysical phenomena. Notably, the Chandrasekhar mass is directly related to the gravitational term as $M_{\rm Ch}\propto G_{\rm eff}^{-3/2}$. Considering the well-established dependence of the SN Ia absolute magnitude on the Chandrasekhar mass, this proportionality can be extended to yield the following relation, 
\begin{equation} \label{eq:Geff}
    M-M_0=\frac{15}{4}\log\left(\frac{G_{\rm eff}}{G_{\rm N}}\right)\,,
\end{equation}
where $M_0$ represents a local value of the SN Ia absolute magnitude associated with the constant $G_{\rm N}$. Equation~\eqref{eq:Geff} indicates that a 10\% increase in the gravitational constant results in a variation of $\Delta M\approx 0.2$, implying that an increase in the gravitational constant implies that SNe Ia are actually brighter. A further discussion can be found in Ref.~\cite{Gaztanaga:2001fh}. 

\begin{figure}
\centering
\includegraphics[width = \columnwidth,trim={0 1cm 0 0.2cm},clip]{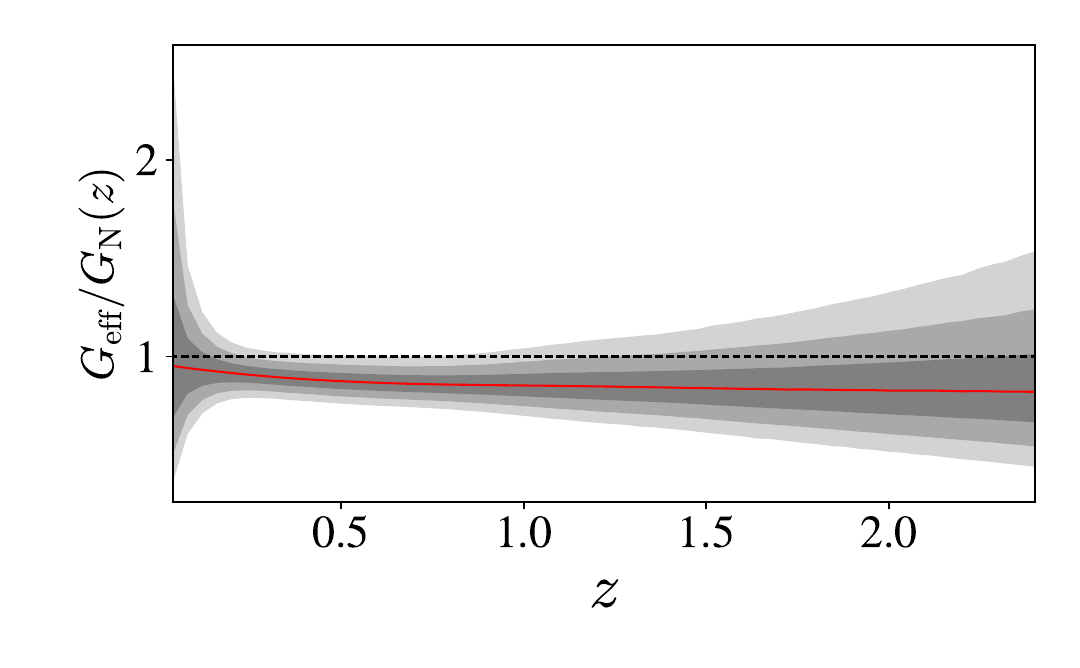}
\caption{Inference of the effective gravitational constant by considering the relation in Eq.  \eqref{eq:GeffoverGN}}. 
\label{fig:Geff}
\end{figure}

Let us now consider a specific theory of gravity wherein the gravitational term is time-dependent $G_{\rm eff}\left(z\right)$, and thereby influencing the variation in the absolute magnitude of SN Ia. Within this framework, the ratio $G_{\rm eff}\left(z\right)/G_{\rm N}$ can be algebraically derived using Eq.~\eqref{eq:Geff}, yielding
\begin{equation} \label{eq:GeffoverGN}
    \frac{G_{\rm eff}}{G_{\rm N}}=10^{\frac{4}{15}\left[M\left(z\right)-M_0\right]}\,.
\end{equation}
Therefore, combining the result obtained for the evolution of the absolute magnitude of SN Ia with Eq.~\eqref{eq:GeffoverGN}, we can derive a time dependence for the ratio $G_{\rm eff}\left(z\right)$. This outcome is illustrated in Fig.~\ref{fig:Geff}, where we have employed the SH0ES value for $M_0$. Similar to the reconstruction of $M(z)$ in comparison with the value obtained through local measurements, it is observed that the best-fit for $G_{\rm eff}\left(z\right)/G_{\rm N}$ is slightly below 1, yet remains compatible with unity within a 3$\sigma$ confidence level. 

\section{Conclusions \label{sec:conclusions}}

We have explored a possible variation in the absolute magnitude of SN Ia by performing  Gaussian process reconstructions  of  independent  cosmological background observables. We  have considered the apparent magnitude $m_b(z)$, the ratio  $D_M(z)/D_H(z)$ and the Hubble rate $H(z)$ provided by Pantheon+SH0ES, SDSS LSS and CC data, respectively.  In order to obtain a robust statistical result, we employed  CPL parameterization as the prior mean function for the Gaussian process regressions and  marginalised over the full set of parameters $\{\sigma,l,\Omega_m,H_0,M,w_0,w_a\}$, including both covariance and cosmological parameters.  This approach  is general enough to yield an unbiased reconstruction for the data considered. Using these observables and the relation \eqref{eq:ME}, we  computed  the mean value, the standard deviation and the covariance across the redshift range of $M(z)$ through Monte Carlo sampling, providing a complete estimate of the redshift evolution of the absolute magnitude of SN Ia and its derivative, shown in Fig. \ref{fig:M_GP}.

We have found no evidence for redshift variation in $M$. Since our results are compatible with a constant value, we have  estimated the redshift average of the absolute magnitude, obtaining$M=-19.456\pm 0.059$. This value is in agreement with both model-dependent and model-independent estimates that use early universe data, but it shows a 3.2$\sigma$ tension with the model-dependent Cepheid variable estimate from SH0ES . Therefore, our results contribute to the well-known discrepancies between late-time and early universe data in the context of the $\Lambda$CDM model.

In the framework of modified gravity theories, these results can be interpreted as evidence for a variation in the effective gravitational term, $G_{\text{eff}}$. When considering the absolute magnitude from Cepheid variables,  the ratio $G_{\text{eff}}/G_{\text{N}}$ reaches a deviation from unity $\sim 3\sigma$ in the redshift range $0.4<z<0.9$.

\section*{Acknowledgements}
The authors thank Savvas Nesseris for useful discussions. RvM is suported by Fundação
de Amparo à Pesquisa do Estado da Bahia (FAPESB)
grant TO APP0039/2023. JA is supported by CNPq grant No. 307683/2022-2 and Funda\c{c}\~ao de Amparo \`a Pesquisa do Estado do Rio de Janeiro (FAPERJ) grant No. 259610 (2021). This work
was developed thanks to the computational support of the National Observatory Data Center (CPDON).

\appendix

\section{Gaussian Process with zero mean function  \label{sec:GP0}}

\begin{figure}
\centering
\includegraphics[width=0.35\textwidth,clip]{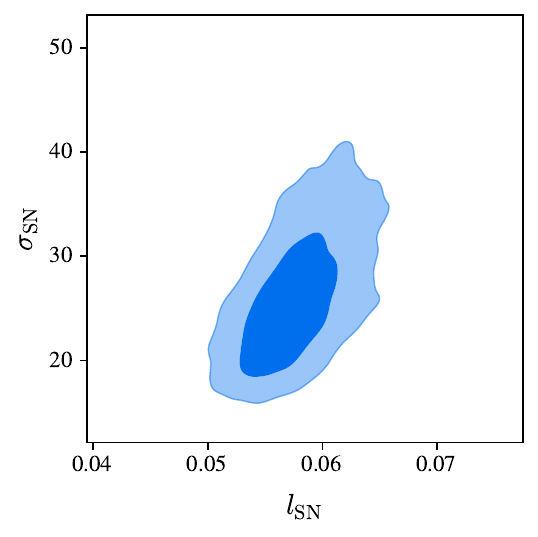}
\caption{Distribution of the hyperparameters of the Gaussian Process for SN Ia Pantheon+SH0ES data assuming a zero prior mean function.}
\label{fig:dist_zero_mean}
\end{figure}

\begin{figure}
\centering
\includegraphics[width = \columnwidth,trim={0 1cm 0 0.2cm},clip]{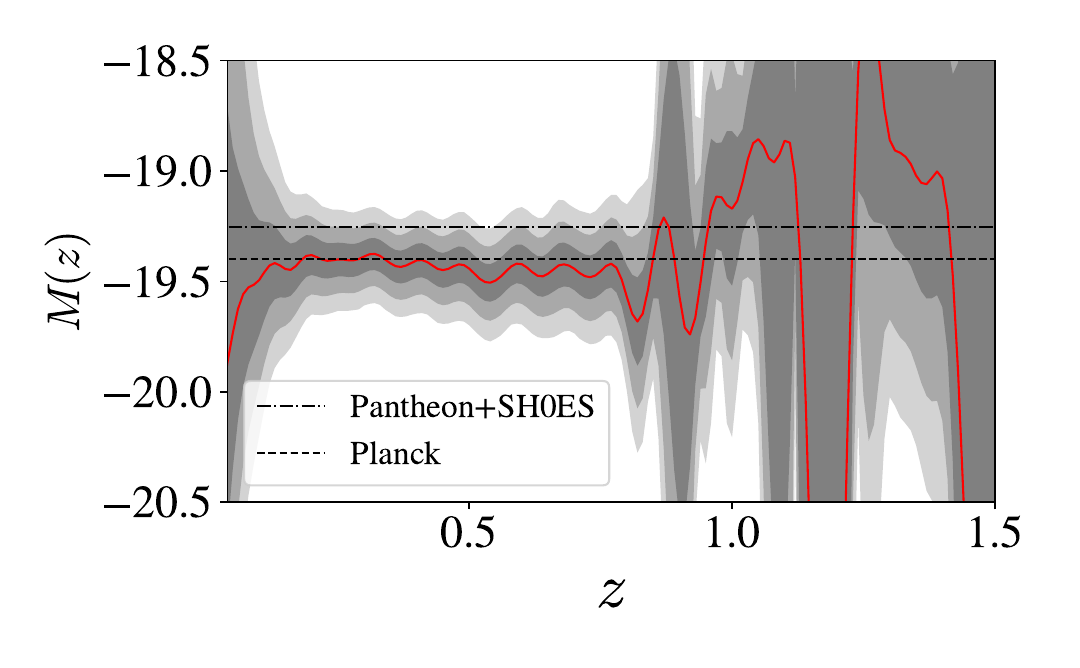}
\includegraphics[width = \columnwidth,trim={0 1cm 0 0.2cm},clip]{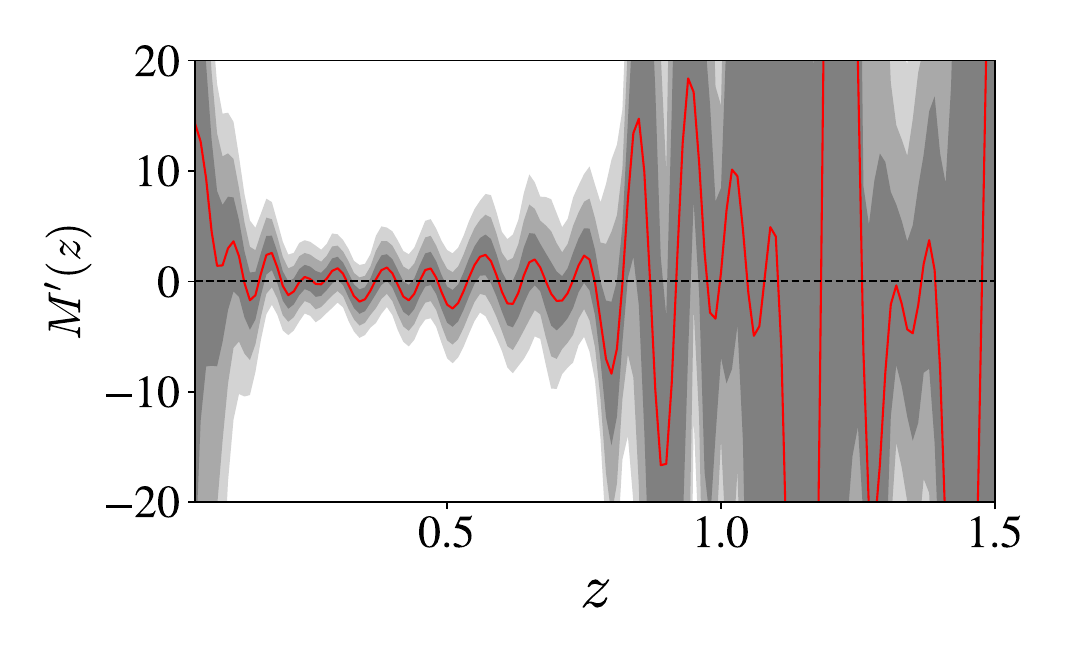}
\caption{\textbf{Left panel:} Absolute magnitude of SN Ia as a function of redshift by considering a zero  function prior in the GP formalism. \textbf{Right panel:} Derivative of absolute magnitude of SN Ia as a function of redshift by considering a zero function prior in the GP formalism.}
\label{fig:M_GP_zero_mean}
\end{figure}

We also explore the reconstructions of all analyzed observables to obtain the redshift evolution of the absolute magnitude by considering a zero prior mean function. In Fig. \ref{fig:dist_zero_mean}, we present the $l-\sigma$ distribution for the SN Ia data. Unlike the case described previously using CPL parameterization as prior function (see Fig. \ref{fig:GP_hyperparams}), in this approach, the data tightly constrain the hyperparameters, leading to a closed contour and a preference for small $l$ values. This leads to a rapidly changing function, causing unphysical wiggles in both the apparent and absolute magnitudes, as shown in Fig. \ref{fig:M_GP_zero_mean}. Naturally, this behavior is also reflected in the derivative of $M$ (see Fig. \ref{fig:M_GP_zero_mean}). These unphysical wiggles justify the use of a CPL parametrization as prior function and,  as it is shown in Ref. \cite{Hwang:2022hla}, this model is enough general to reconstruct the apparent magnitude observable without biased results.


\bibliography{biblio}

\end{document}